\documentclass[conference]{IEEEtran}
\usepackage[utf8]{inputenc}   
\usepackage{cite}             
\usepackage[pdftex]{graphicx} 
\usepackage{array}
\usepackage{url}
\usepackage{microtype}
\usepackage[nameinlink]{cleveref}
\usepackage{tikz}

\newcommand\copyrighttext{%
  \footnotesize \textcopyright 2018 IEEE. Personal use of this material is permitted. Permission from IEEE must be obtained for all other uses, in any current or future media, including reprinting/republishing this material for advertising or promotional purposes, creating new collective works, for resale or redistribution to servers or lists, or reuse of any copyrighted component of this work in other works. DOI: \url{https://doi.org/10.1109/FIE.2018.8658840}.}
\newcommand\copyrightnotice{%
\begin{tikzpicture}[remember picture,overlay]
\node[anchor=south,yshift=10pt] at (current page.south) {\fbox{\parbox{\dimexpr\textwidth-\fboxsep-\fboxrule\relax}{\copyrighttext}}};
\end{tikzpicture}%
}

\begin{document}

\title{Gathering Insights from Teenagers' Hacking Experience with Authentic Cybersecurity Tools}

\author{
\IEEEauthorblockN{Valdemar Švábenský}
\IEEEauthorblockA{Faculty of Informatics\\
Institute of Computer Science\\
Masaryk University\\
Brno, Czech Republic\\
svabensky@ics.muni.cz}
\and
\IEEEauthorblockN{Jan Vykopal}
\IEEEauthorblockA{Institute of Computer Science\\
Masaryk University\\
Brno, Czech Republic\\
vykopal@ics.muni.cz}
}

\maketitle
\copyrightnotice

\begin{abstract}
This Work-In-Progress Paper for the Innovative Practice Category presents a novel experiment in active learning of cybersecurity. We introduced a new workshop on hacking for an existing science-popularizing program at our university. The workshop participants, 28 teenagers, played a cybersecurity game designed for training undergraduates and professionals in penetration testing. Unlike in learning environments that are simplified for young learners, the game features a realistic virtual network infrastructure. This allows exploring security tools in an authentic scenario, which is complemented by a background story. Our research aim is to examine how young players approach using cybersecurity tools by interacting with the professional game. A preliminary analysis of the game session showed several challenges that the workshop participants faced. Nevertheless, they reported learning about security tools and exploits, and 61\% of them reported wanting to learn more about cybersecurity after the workshop. Our results support the notion that young learners should be allowed more hands-on experience with security topics, both in formal education and informal extracurricular events.
\end{abstract}

\section{Introduction}

Learning cybersecurity is no longer solely the domain of university students and adult professionals. As security topics grow in importance, they also emerge in K-12 education. One of the main reasons for this is the rapidly rising demand for cybersecurity workforce and related technical professions. As a result of this demand, many educational institutions popularize science, technology, engineering, and mathematics (STEM) by introducing active learning methods and extracurricular activities. Their goal is to engage young students and motivate them to pursue technical careers.

Our university is no exception, as it popularizes STEM among youth by regularly hosting a public event called ``Junior University\footnote{\url{http://mjuni.cz/}}.'' For this event, we introduced an innovative hands-on workshop about hacking. Our goal was to raise awareness of the cybersecurity field via engaging and authentic experience with tools used in professional practice. By letting the learners use these tools, we wanted to show them real-life aspects of hacking, as opposed to its popular but inaccurate display in movies and video games.

To achieve our goal, we employed the format of a Capture the Flag (CTF) game, which is often used to practice cybersecurity skills. While there are different types of CTF games, we focus on an Attack-only CTF, in which the player receives offensive security tasks. Completing each task yields a unique textual \textit{flag}. This flag is used to confirm or deny the solution and automatically award points to the player. We deploy these games in the KYPO cyber range -- a virtual environment that allows the player to exercise cybersecurity tools in a realistic network setting~\cite{kypo}. At the same time, the player's actions are isolated from the outside world and cannot have any negative consequences in reality, which allows free experimentation.

Our research aim is to examine and understand how young players interact with the professional cybersecurity game and the network security tools. By analyzing the game logs, we want to gain insights into which tasks the learners achieved and which were problematic and why. As a result, we can substantially improve upon our prior practice by redesigning the game tasks and scaffolding. This will enhance the learning experience of future young players. Moreover, since other security games feature similar player interaction, our contributions to cybersecurity education will be relevant also for other authors in perfecting their games.

\section{Related Work}

There are several notable efforts in cybersecurity education for K-12 students. Related CTF games include PicoCTF\cite{Chapman}, a web-based computer security competition for high school students focused on offensive skills. CTF Unplugged\cite{Ford} is an offline competition to motivate and teach high school students with little or no technical knowledge. There has been research on how CTF games with an engaging storyline improve students' performance\cite{chothia}. Other relevant games are Netsim\cite{Atwater}, a web game that teaches network attacks to high schoolers, and Ctrl-Alt-Hack\cite{Denning}, a card game for teaching security. Next, materials for teaching authentication in high schools have been developed and tested\cite{Stobert}. Finally, after-school cybersecurity camps are popular. CyberPatriot~\cite{Dunn} is a competition for increasing K-12 students' interest in security. NSA/NSF GenCyber program\cite{Ladabouche} raises cybersecurity awareness for K-12 students. Our workshop adds to all these efforts by letting the young participants play a professional cybersecurity game in a realistic, non-simplified environment.

\section{Methodology and Experiment Setup}

This section describes the design of our workshop and the experiment performed during it to address our research aim. The learning objectives of the workshop were that the participants:
\begin{itemize}
\item Gain awareness about cybercrime;
\item Learn about different types of cyber attacks;
\item Practice using network security tools;
\item Know the principles of creating secure passwords; and
\item Understand basic ethical and legal aspects of hacking.
\end{itemize}

\subsection{Participants}
28 self-selected teenagers (21 boys, 7 girls) voluntarily signed up for our workshop at the Junior University event. They chose our activity among five others offered. Their only motivation for attending was their interest, as they did not receive any incentives for taking part. The participants' age ranged from 13 years (11$\times$), through 14 years (16$\times$), to 15 years (1$\times$); therefore, they were enrolled by their parents.

\subsection{Structure and content of the workshop}
Two Ph.D. students, one of which is the author of this paper, facilitated the workshop along with one undergraduate. \Cref{table:schedule} shows the schedule. We started with a lecture about hacking, cyber attack stages\footnote{Reconnaissance, scanning, gaining access, and exploiting a vulnerability} explained on examples, and methods of password attacks. Although we have experience mainly with university-level teaching, we prepared the lecture in an engaging and simplified way. We considered the audience's age and included informative videos and illustrations. The lecture was followed by an interactive quiz on Kahoot\footnote{\url{https://kahoot.com}} to practice the discussed concepts on 10 multiple choice questions. Next, the attendees proceeded to the game, described in the \cref{subsec:game}.

After the learners practically tried out real network security tools, we instructed them about ethical and legal aspects of hacking, the importance of understanding cyber attacks in setting up effective defenses, and ``white hat'' hacking for beneficial purposes. Moreover, all participants received a handout with a summary of tips for creating secure passwords. At the very end, we asked all participants to complete a survey described in the \cref{subsec:survey}.

\begin{table}[!t]
\renewcommand{\arraystretch}{1.2}
\caption{Schedule of our hacking workshop}
\label{table:schedule}
\centering
\begin{tabular}{|l|p{2.4cm}|l|l|p{1.4cm}|}
\hline
\textbf{Activity}   & Introductory lecture (including questions) & Kahoot & Game & Conclusion, survey \\ \hline
\textbf{Time [min]} & 20 & 10 & 90 & 15 \\ \hline
\end{tabular}
\end{table}

\subsection{Selected cybersecurity game}\label{subsec:game}
The attendees played a cybersecurity game called \textit{Photo Hunter}, which was created by the students of our Cyber Attack Simulation course\cite{ITICSE}. The story of the game features three characters: Sarah, a popular celebrity; Paul, a paparazzi; and the player, who is given the role of a renowned hacker. The paparazzi blackmails the celebrity, as he claims to possess incriminating photographs of her and threatens to publish them unless he receives a large sum of money. Sarah is not sure whether the photos even exist, and so asks the player, her friend from high school, for help.

The player begins the game by receiving control of a single attacker machine. The machine runs Kali Linux\footnote{\url{https://www.kali.org}} operating system in a realistic virtual environment emulated by the KYPO cyber range. The learning objective of the player is to exercise four stages of a cyber attack in four consecutive levels. First, the player has to examine headers of the blackmailing e-mail sent from the paparazzi's web server, find the IP address of the server, and perform an \texttt{nmap} scan to discover its open ports. Second, the player connects to the server, learns that it hosts a WordPress website, and performs a vulnerability scan with the \texttt{wpscan} tool. The scan reveals an SQL injection vulnerability, which is exploited in the third level using \texttt{sqlmap}, leading to the discovery of a photo storage server. Finally, the player performs a dictionary attack on the server using \texttt{medusa}. However, gaining access leads to a surprising conclusion: the paparazzi's server stores only funny pictures of pigs.

The game was played individually, although we encouraged collaboration if it occurred later in the game. Still, the three facilitators provided most individual help to the participants. Since the game was tested in practice several times, we knew the players' common pitfalls and aided them effectively. The players also asked for hints or even complete solutions directly in the game interface without any assistance of facilitators.

We logged the game actions and the game score of each participant as in our previous research~\cite{SIGCSE}. The game events describe the player's interaction with the game interface, namely: starting and ending the game or each level, submitting incorrect flags and their content, taking hints, skipping a level, and displaying a solution. Each game event contains a timestamp and a unique ID of the player, which allowed us to aggregate the logged actions. Each ID was assigned randomly, and we did not associate any personally identifiable information with it to protect the privacy of the attendees.

\subsection{After-game survey}\label{subsec:survey}
We designed a 10-item survey to understand the participants' experience and motivation after the workshop. We followed an extensive literature search of best practices in survey design~\cite{surveys}. At the same time, we kept the questionnaire brief in order not to bore the young participants and not to discourage them from completing it. The survey was optional, anonymous, and administered online via Google Forms. Although the whole text was in Czech language to ensure that the participants understand the survey accurately, its translation follows:
\begin{enumerate}
\item How much are you satisfied or dissatisfied with the workshop? (Not at all satisfied / Slightly satisfied / Moderately satisfied / Very satisfied / Extremely satisfied)
\item What did you like or dislike the most?
\item Were you interested in computer security before today's workshop? (Yes / No)
\item Do you want to learn more about computer security after today’s workshop? (Yes / No)
\item If you answered Yes to the previous question, what specifically would you like to learn?
\item Are you considering to enroll in a high school (or even a university) focused on computer science? (Yes / No)
\item What was your 7-digit game ID?
\item How easy or difficult was the game for you? (Very easy / Easy / Medium, balanced / Difficult / Very difficult)
\item What are your most important learning experiences from today? What will you definitely remember?
\item Do you want to leave a comment for the organizers? Now you have the chance!
\end{enumerate}

\begin{figure*}[!t]
\centering
\includegraphics[width=\textwidth]{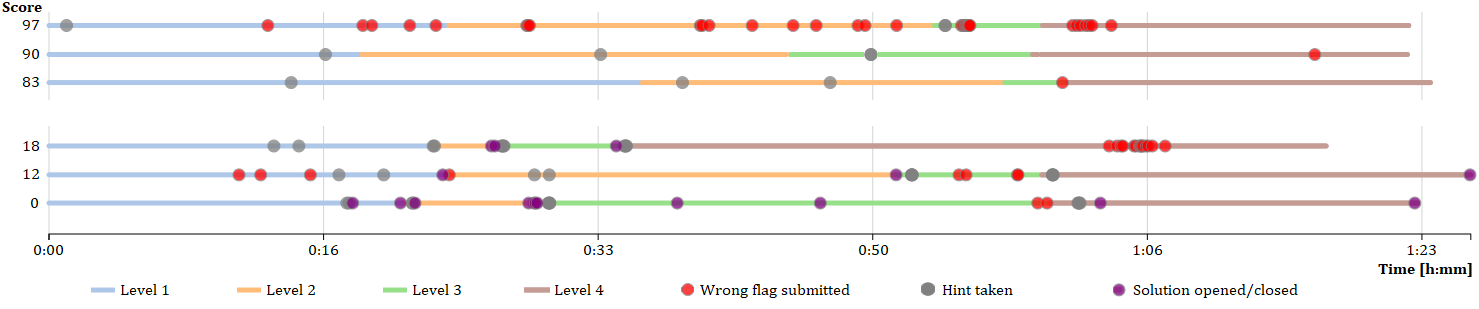}
\caption{Game events of three best and three worst scoring players distributed over time; each timeline depicts important game events of one player}
\label{figure:events}
\end{figure*}

\section{Preliminary Results and Discussion}

\subsection{After-game survey}
We evaluated the survey responses of 27 participants. The remaining one did not respond for unknown reasons. The participants reported high satisfaction with the workshop: 3~learners were extremely satisfied, 16 very satisfied, 4 moderately satisfied, and 4 slightly satisfied. The most prominent likes include: ``the workshop was well-prepared and fun'', ``I liked trying new things'', and ``I'm thankful for the opportunity to try the game''. The players appreciated learning by doing, using Linux and hacking tools, and willingness of the tutors. The only reported dislikes were: difficulty to understand some tasks in English, minor bugs in the cyber range, score penalization for taking hints, and difficulty of the game. 7 players found it extremely difficult, 15 difficult, and only 7 medium.

13 learners (46\%) were interested in security before the workshop, and 17 (61\%) wanted to learn more afterward. The students were curious namely about programming, more network security tools and how do they work ``under the hood'', password cracking, and how to secure their computers and internet banking. Among all the learners, 13 (46\%) considered enrolling in a school focused on computer science.

The learners' self-reported most important learning experiences include learning about attack types and how to perform them, Kali Linux commands (\texttt{nmap} and \texttt{medusa} were mentioned explicitly), and how SQL injection works. The learners also understood the legal implications of hacking. Some attendees shared their experiences in more detail:
\begin{itemize}
\item ``If I'm nice and report security holes, I can earn money.''
\item ``I was surprised that people use really dumb passwords. I should use better passwords.''
\item ``I learned that hacking is illegal unless it’s agreed upon.'' (referring to penetration testing)
\item ``Real computer science is really difficult, compared to what we do at elementary school.''
\item ``I can search garbage cans to find documents with sensitive information.''
\end{itemize}

\subsection{In-game data}

The game score could range from 0 to 100 points. In our sample, the minimum was 0, the maximum 97, the average 54, and the median 55 points. Points were awarded for submitting a correct flag in each level and deducted for taking hints or displaying solutions. We selected three best and three worst scoring players with the aim to compare their game strategy. \Cref{figure:events} visualizes their game events. (The visualization was described in our previous paper\cite{SIGCSE} and has been improved since then.) As displaying a solution causes scoring penalty, the figure shows that the lowest scoring players viewed most or all the solutions, while the top scoring players did not display any solution. Perhaps surprisingly, the top scoring player attempted to submit a lot of incorrect flags throughout the game. This player generated a burst of 14 wrong flags within 3 minutes in the last level, which shows guessing. A final remark about these selected players is that cross-referencing the game events with the survey showed the first three players were all interested in cybersecurity before the workshop. On the contrary, the last three players reported not having a previous interest.

Our game logs confirm the observation of the facilitators that some learners had trouble understanding the game mechanics or user interface. The interface consisted of two windows: one to control the Linux attacker machine, another to submit flags and display the game's rules, objectives, and hints. However, at the beginning of the game, 9 players entered their virtual machine login credentials into the flag input form. In general, flag format and its submission were puzzling for several players. Although the flag for the first level was a network port number, 13 players entered strings that were not port numbers, which indicates a lack of knowledge in the area. While the majority of players cleared their misunderstanding of flag format after the first level, 4 of them still submitted a text string as a flag in the second level that again required a number.

We will now comment on game hints and solutions. Within the first 10 seconds from the start, 3 players took a hint. This indicates the game was likely difficult for these players, as they requested help without even reading the assignment. In general, the facilitators observed that many students got stuck because they did not read the instructions properly. However, throughout the whole game, only 7 players exercised their option to view a solution to a level. Others refused to display the solution or even take too many hints, as this induced a score penalty. Although there was no competition and the displayed game points had virtually no significance, those more competitive took their score very seriously. Thus, to encourage taking hints and support learning, we suggest removing the score penalization from the game when not using it for competitions.

\subsection{Observations of the facilitators}
We now present the mutual conclusions of the three facilitators. The game was challenging for most attendees, which they reported in the survey. Apart from the struggles reflected in the game data, the most common difficulty was using a command-line interface. Half of the learners had never before used a Linux operating system and needed, first and foremost, explaining the workflow with the Terminal and command syntax.

Despite the issues we encountered, we could see that most players were fascinated by the game. To improve their experience further and have more time for individual help, we could use at least twice the number of facilitators on the class of the same size (that is, one facilitator for approximately five learners). Although we encouraged cooperation between neighboring peers, too, some of them refused, perhaps due to not knowing each other and being shy.

\section{Conclusions}

So far, we succeeded in running the event, gathering the in-game and survey data, and preliminary analyzing them. As this work is still in progress, we will thoroughly analyze the game events to better understand players' interaction with the game. Our primary motivation is being able to improve the game and learning experiences of the players. Since we can connect questionnaire results to the game events based on the player ID, we will also examine the relationship between players' actions and feedback. For our further future work, we plan to collect Photo Hunter game events from other demographic groups, such as students pursuing a CS degree, and compare them with each other. We are curious to see if there are any differences in the interactions or strategy patterns of the players.

One of the key takeaways from this article is our recommendation for cybersecurity instructors to organize similar popularizing activities. Even though our game was designed primarily for university students and adults, the positive feedback from our players showed that the game has beneficial influence also on younger learners. The participants reported learning about expert tools, mechanics behind particular exploits, and legal aspects of hacking. Although the game was challenging for 79\% of learners, they appreciated working with real security tools, and 68\% reported satisfaction with the workshop.

\section*{Acknowledgment}
This research was supported by the Security Research Programme of the Czech Republic 2015--2020 (BV III/1--VS) granted by the Ministry of the Interior of the Czech Republic under No.~VI20162019014 -- Simulation, detection, and mitigation of cyber threats endangering critical infrastructure. We thank the students and tutors of our Cyber Attack Simulation course for designing the Photo Hunter game.

\bibliographystyle{IEEEtran}
\bibliography{fie2018}

\begin{thebibliography}{10}
\providecommand{\url}[1]{#1}
\csname url@samestyle\endcsname
\providecommand{\newblock}{\relax}
\providecommand{\bibinfo}[2]{#2}
\providecommand{\BIBentrySTDinterwordspacing}{\spaceskip=0pt\relax}
\providecommand{\BIBentryALTinterwordstretchfactor}{4}
\providecommand{\BIBentryALTinterwordspacing}{\spaceskip=\fontdimen2\font plus
\BIBentryALTinterwordstretchfactor\fontdimen3\font minus
  \fontdimen4\font\relax}
\providecommand{\BIBforeignlanguage}[2]{{%
\expandafter\ifx\csname l@#1\endcsname\relax
\typeout{** WARNING: IEEEtran.bst: No hyphenation pattern has been}%
\typeout{** loaded for the language `#1'. Using the pattern for}%
\typeout{** the default language instead.}%
\else
\language=\csname l@#1\endcsname
\fi
#2}}
\providecommand{\BIBdecl}{\relax}
\BIBdecl

\bibitem{kypo}
J.~Vykopal, R.~Oslejsek, P.~Celeda, M.~Vizvary, and D.~Tovarnak, ``{KYPO Cyber
  Range: Design and Use Cases},'' in \emph{Proceedings of the 12th
  International Conference on Software Technologies -- Volume 1: ICSOFT},
  INSTICC.\hskip 1em plus 0.5em minus 0.4em\relax SciTePress, 2017, pp.
  310--321.

\bibitem{Chapman}
\BIBentryALTinterwordspacing
P.~Chapman, J.~Burket, and D.~Brumley, ``{PicoCTF: A Game-Based Computer
  Security Competition for High School Students},'' in \emph{2014 {USENIX}
  Summit on Gaming, Games, and Gamification in Security Education (3GSE
  14)}.\hskip 1em plus 0.5em minus 0.4em\relax San Diego, CA: {USENIX}
  Association, 2014. [Online]. Available:
  \url{https://www.usenix.org/conference/3gse14/summit-program/presentation/chapman}
\BIBentrySTDinterwordspacing

\bibitem{Ford}
\BIBentryALTinterwordspacing
V.~Ford, A.~Siraj, A.~Haynes, and E.~Brown, ``{Capture the Flag Unplugged: An
  Offline Cyber Competition},'' in \emph{Proceedings of the 2017 ACM SIGCSE
  Technical Symposium on Computer Science Education}, ser. SIGCSE '17.\hskip
  1em plus 0.5em minus 0.4em\relax New York, NY, USA: ACM, 2017, pp. 225--230.
  [Online]. Available: \url{http://doi.acm.org/10.1145/3017680.3017783}
\BIBentrySTDinterwordspacing

\bibitem{chothia}
\BIBentryALTinterwordspacing
T.~Chothia, S.~Holdcroft, A.-I. Radu, and R.~J. Thomas, ``{Jail, Hero or Drug
  Lord? Turning a Cyber Security Course Into an 11 Week Choose Your Own
  Adventure Story},'' in \emph{2017 {USENIX} Workshop on Advances in Security
  Education}, Vancouver, BC, 2017. [Online]. Available:
  \url{https://www.usenix.org/conference/ase17/workshop-program/presentation/chothia}
\BIBentrySTDinterwordspacing

\bibitem{Atwater}
\BIBentryALTinterwordspacing
E.~Atwater, C.~Bocovich, U.~Hengartner, and I.~Goldberg, ``{Live Lesson:
  Netsim: Network simulation and hacking for high schoolers},'' in \emph{2017
  {USENIX} Workshop on Advances in Security Education ({ASE} 17)}.\hskip 1em
  plus 0.5em minus 0.4em\relax Vancouver, BC: {USENIX} Association, 2017.
  [Online]. Available:
  \url{https://www.usenix.org/conference/ase17/workshop-program/presentation/atwater}
\BIBentrySTDinterwordspacing

\bibitem{Denning}
\BIBentryALTinterwordspacing
T.~Denning, A.~Shostack, and T.~Kohno, ``{Practical Lessons from Creating the
  Control-Alt-Hack Card Game and Research Challenges for Games In Education and
  Research},'' in \emph{2014 {USENIX} Summit on Gaming, Games, and Gamification
  in Security Education}, San Diego, CA, 2014. [Online]. Available:
  \url{https://www.usenix.org/conference/3gse14/summit-program/presentation/denning}
\BIBentrySTDinterwordspacing

\bibitem{Stobert}
\BIBentryALTinterwordspacing
E.~Stobert, E.~Cavar, L.~Malisa, and D.~Sommer, ``{Teaching Authentication in
  High Schools: Challenges and Lessons Learned},'' in \emph{2017 {USENIX}
  Workshop on Advances in Security Education ({ASE} 17)}.\hskip 1em plus 0.5em
  minus 0.4em\relax Vancouver, BC: {USENIX} Association, 2017. [Online].
  Available:
  \url{https://www.usenix.org/conference/ase17/workshop-program/presentation/stobert}
\BIBentrySTDinterwordspacing

\bibitem{Dunn}
\BIBentryALTinterwordspacing
M.~H. Dunn and L.~D. Merkle, ``{Assessing the Impact of a National
  Cybersecurity Competition on Students' Career Interests},'' in
  \emph{Proceedings of the 49th ACM Technical Symposium on Computer Science
  Education}, ser. SIGCSE '18.\hskip 1em plus 0.5em minus 0.4em\relax New York,
  NY, USA: ACM, 2018, pp. 62--67. [Online]. Available:
  \url{http://doi.acm.org/10.1145/3159450.3159462}
\BIBentrySTDinterwordspacing

\bibitem{Ladabouche}
T.~Ladabouche and S.~LaFountain, ``{GenCyber: Inspiring the Next Generation of
  Cyber Stars},'' \emph{IEEE Security \& Privacy}, vol.~14, no.~5, pp. 84--86,
  2016.

\bibitem{ITICSE}
\BIBentryALTinterwordspacing
V.~\v{S}v\'{a}bensk\'{y}, J.~Vykopal, M.~Cermak, and M.~La\v{s}tovi\v{c}ka,
  ``{Enhancing Cybersecurity Skills by Creating Serious Games},'' in
  \emph{Proceedings of the 23rd Annual ACM Conference on Innovation and
  Technology in Computer Science Education}, ser. ITiCSE '18.\hskip 1em plus
  0.5em minus 0.4em\relax ACM, 2018. [Online]. Available:
  \url{http://doi.acm.org/10.1145/3197091.3197123}
\BIBentrySTDinterwordspacing

\bibitem{SIGCSE}
\BIBentryALTinterwordspacing
V.~\v{S}v\'{a}bensk\'{y} and J.~Vykopal, ``{Challenges Arising from
  Prerequisite Testing in Cybersecurity Games},'' in \emph{Proceedings of the
  49th ACM Technical Symposium on Computer Science Education}, ser. SIGCSE
  '18.\hskip 1em plus 0.5em minus 0.4em\relax ACM, 2018, pp. 56--61. [Online].
  Available: \url{http://doi.acm.org/10.1145/3159450.3159454}
\BIBentrySTDinterwordspacing

\bibitem{surveys}
\BIBentryALTinterwordspacing
E.~M. Redmiles, Y.~Acar, S.~Fahl, and M.~L. Mazurek, ``{A Summary of Survey
  Methodology Best Practices for Security and Privacy Researchers},'' Tech.
  Rep., 2017. [Online]. Available: \url{http://hdl.handle.net/1903/19227}
\BIBentrySTDinterwordspacing

\end{thebibliography}
\end{document}